# An energy gap in the spectrum of atomic excitations $He^4$ systems.

## A.I. Karasevskii

G.V. Kurdymov Institute for Metal Physics of Ukrainian Academy of Sciences

36 Vernadsky bl., Kiev, 03142, Ukraine

E-mail: akaras@imp.kiev.ua

**Abstract**

It is shown that atoms in the $He^4$ system can be treated as quantum particles localized in potential wells, created by atomic potentials of neighboring atoms. As a result, the state of atoms in the liquid $He^4$ is characterized by wave functions and the discrete energy spectrum resulting in formation of s – and p- zones corresponding to the ground and excited states of helium atoms, respectively, separated by a gap. The width of the gap in $He^4$ system equals ~8.5K at T=0. Formation of the gap in the energy spectrum of atomic excitations in helium systems allows us to draw the analogy between the physical mechanisms of superfluidity and classical superconductivity.

**Keywords:** liquid helium, energy gap, superfluidity

## 1. Introduction

In classical crystals and liquids, atoms are localized in parabolic potential wells and are involved in collective quantum and thermal motion. In contrast, in low-density helium systems, atoms are localized in essentially non-parabolic potential wells, and collective quantum motion of atoms is characterized by only shot-range correlations between movements of neighboring atoms, arising from strong interatomic repulsion at short distances. Thus, at $T = 0$, the quantum motion of a helium atom in an effective potential well is, to a large extent, individual, with only pairwise interatomic correlations taken into account. In such a potential well, a helium atom, being a quantum particle, may occupy either the ground $s$- state or an excited p-state with different energies [1,2]. As was shown before [1, 2], in the case of $1D$ $He^4$ crystal, the distance $\Delta\varepsilon_p$ between the ground energy level and the first excited one depends strongly on external pressure and is about 5-7 K at normal conditions.

In this paper, we show that in 3D $He^4$ crystals or liquid helium, quantum properties of helium atoms are manifested in formation of a discrete spectrum of atomic excitations, the collective nature of which leads to the formation of s and p zone ground and excited states of helium atoms, separated by a gap. In this connection, excitation of an atom in the ground state requires more energy to overcome the energy gap. Such structure of the energy spectrum of atomic excitations in the system allows us to draw an analogy between the physical mechanisms of superfluidity and superconductivity.

Really, at T=0, all the helium atoms are in the ground state, and they would not be excited at low fluid velocities, i.e. the fluid flow will occur without viscous friction, showing the superfluid nature. If the flow velocity exceeds a certain critical value of the excitation energy of the atoms, and the excitation energy of atoms exceeds the band gap, the superfluidity properties disappear.

Transition of a part of the atoms into the $p$ band can also occur with increasing temperature, while a part of atoms still remains in the ground state, retaining superfluid properties.

The effects of electric induction [3] and the resonant absorption of microwaves [4] which had been observed in superfluid $He^4$ may be considered as an indirect proof of the existence of the quantum transitions in superfluid helium.

These effects can be explained by violation of the adiabatic conditions in the electron-nuclear subsystem of a $He^4$ atom.

## 2. The ground state of helium atoms in a medium

To describe quantum states of an atom in a crystal or a liquid, we will start from the Hamiltonian

$$H = -\frac{\hbar^2}{2m}\sum_i \nabla_i^2 + \frac{1}{2}\sum_{i,j} u(r_{ij}), \qquad (1)$$

where $r_{ij} = |\mathbf{r}_i - \mathbf{r}_j|$ - is a distance between atoms $i$ and $j$, $u(r_{ij})$ is a pairwise interatomic potential, which is approximated by the Morse potential

$$u(r_{ij}) = A\left(e^{-2\alpha(r_{ij}-R_0)} - 2e^{-\alpha(r_{ij}-R_0)}\right). \qquad (2)$$

The values of the parameters $A, \alpha$ and $R_0$ were calculated so that (2) gave the best fit of the Aziz potential of interaction of $He$ atoms [5,6] ( $A = 10.8K$, $\alpha = 2.063 A^{0-1}$ and $R_0 = 2.97 A^0$ ).

In this paper, the calculation of the energy spectrum of the helium atom in the medium is restricted with the quasicrystalline approximation, i.e. we assume that the short-range order in the medium has quasicrystalline structure. The spatial distribution of the potential energy $U_0(r_i)$ near the site $i$ of the bcc lattice is obtained by summation of (2) over the first and second coordination spheres of site $i$ whose radii are $R$ and $2R/\sqrt{3}$, respectively.

### 2a. The bcc short-range order

In the case of the short-range order with a bcc structure, summation over sites of the first and second coordination spheres (the origin of the coordinates is chosen at $r=0$) gives:

$$U_0(t) = 8A\left[\left(1+\frac{3q^2}{4}\right)e^{-2b}\cosh\left(\frac{2t}{\sqrt{3}}\right) - 2\left(1+\frac{3q}{4}\right)e^{-b}\cosh\left(\frac{t}{\sqrt{3}}\right)\right], \qquad (3)$$

where the value of

$$q = e^{-\frac{2-\sqrt{3}}{\sqrt{3}}\alpha R},$$

takes into account a contribution of atoms of the second coordination sphere to $U_0$, $b = \alpha(R-R_0)$ determines the quantum lattice expansion, $t = \alpha r$ is a displacement of atom $i$ from its site.

With the potential (3), the Hamiltonian (1) splits into a sum of independent Hamiltonian, and the Schrödinger equation for the radial part of the ground state wave function,

$$\psi_0(r) = \frac{\chi(r)}{r} \qquad (4)$$

is written as

$$\chi''(t) + \left(k^2 - s^2\cosh^2 t + s^2 P\cosh t + \frac{s^2}{2}\right)\chi(t) = 0, \qquad (5)$$

Where

$$P = \frac{1+\frac{3}{4}q}{1+\frac{3}{4}q^2}e^b,$$

$$s = \frac{4\sqrt{6}}{\Lambda}e^{-b}\left(1+\frac{3}{4}q^2\right)^{1/2}, \qquad (6)$$

$$k^2 = 6\frac{E}{A}\frac{1}{\Lambda^2},$$

$E$ is the system energy, and

$$\Lambda = \frac{\hbar\alpha}{\sqrt{mA}}.$$

is the de Boer parameter for the Morse potential ($\Lambda = 2.17$ for the $He^4$ atom), $m$ is the atomic mass.

A solution of (5), vanishing at $t \to \infty$, is

$$\chi(t) \sim e^{-s\cosh t}.$$

After substitution

$$\chi(\eta) = e^{-s(1+\eta)} \eta^{1/2} \mu(\eta) \qquad (7)$$

where $\eta = 2\sinh^2 \frac{t}{2}$, equation (5) takes on the form

$$\eta(\eta+2)\mu''(\eta) + [3+2\eta-2s(2+\eta)\eta]\mu'(\eta) + Q(\eta)\mu(\eta) = 0, \qquad (8)$$

where

$$Q(\eta) = \frac{1}{4} + k^2 - \frac{s^2}{2} - s(3+2\eta) + s^2 P(1+\eta).$$

The (8) is a differential equation of the second order containing a regular singular point ($\eta \to 0$) and an irregular one ($\eta \to \infty$). Taking into account (8), one can conclude that the regular singular point $\eta \to 0$ determines the solution of (8) (see [7]). Let us assume the solution of the equation (8) near the regular singular point to be of the form of a generalized polynomial:

$$\mu(\eta) = 1 + \gamma \eta^\beta. \qquad (9)$$

Substituting (9) into (8) and equating the coefficients of corresponding powers of $\eta$, we find parameters $\gamma, \beta$ and calculate the energy of the system:

$$\beta = \frac{1}{2} Ps - 1, \quad \gamma \approx 0.65. \qquad (10)$$

Energy of the system is:

$$\varepsilon = \frac{E}{A} = \frac{\Lambda^2}{6}\left(3s + \frac{s^2}{2} - Ps^2 - 3\beta\gamma - \frac{1}{4}\right), \qquad (11)$$

$v = 4R^3 / 3\sqrt{3}$ is volume per atom.

Thus, the energy of the crystal is a function of the variational parameter $b$ determining the quantum lattice expansion. From (4), (7) and (9), we get an explicit expression for the ground state wave function of an atom localized at the site $r = 0$

$$\psi_0(r) = C_0 \frac{\sinh\left(\frac{\alpha r}{2\sqrt{3}}\right)}{r} e^{-s\cosh\left(\frac{\alpha r}{\sqrt{3}}\right)} \left[1 + \gamma \sinh\left(\frac{\alpha r}{2\sqrt{3}}\right)\right]^\beta, \qquad (12)$$

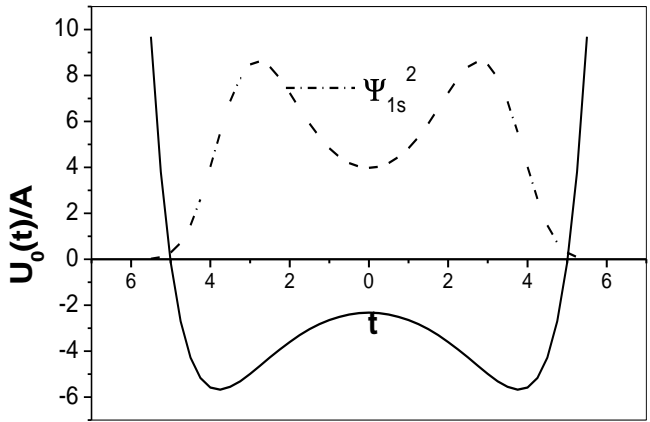

Fig.1a.

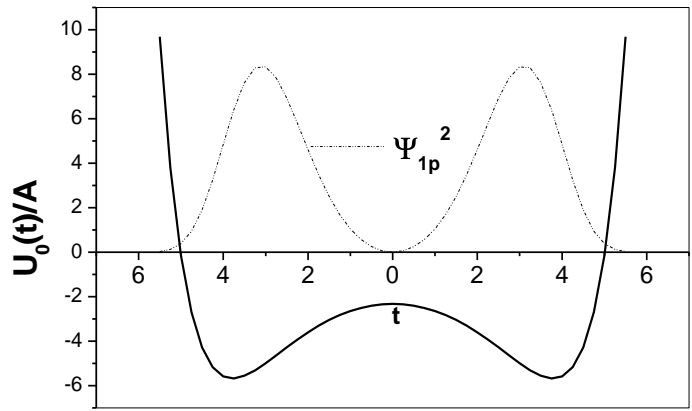

Fig. 1b

Fig.1. Distribution of the ground-state wavefunction (Fig.1a) and excited-state wavefunction (Fig.1b) of the He atom in the potential well.

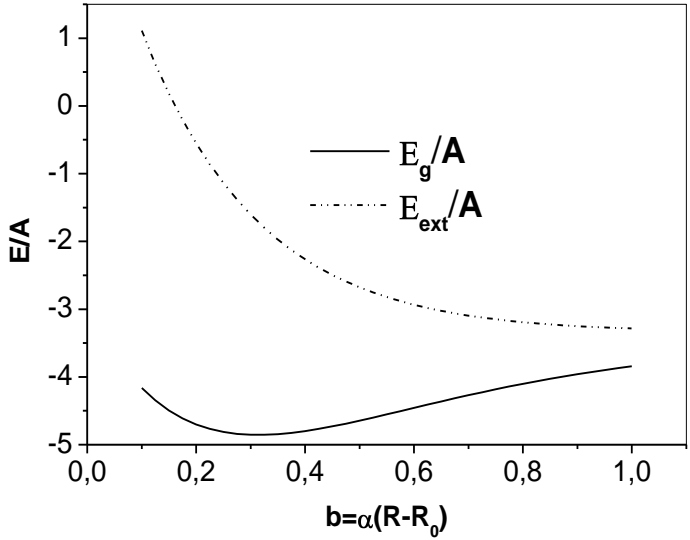

Fig.2. Dependence ground-state ($E_g$) and excited-state ($E_{ext}$) energy on interatomic expansion.

## 2. The excited state of a helium atom in the medium

The basic peculiarity of the hellium crystals and liquids is that their atoms should be considered as quantum particles moving in potential wells created by interaction with neighbouring atoms. A localized state of a quantum particle in such a potential well is characterized by a set of energy levels and corresponding wave functions. In a spherical symmetric potential well, the first excited state of an atom is the p-state, whose wave function vanishes at the center of the well and reaches its maximum value near the boundary of the atomic cell. Such probability of spatial distribution of the particle in p-state is well appropriate to the spatial distribution of the potential energy in atomic cell of the helium medium (Fig.1).

Thus one may propose that, even though the kinetic energy in the p-state is greater than that in the ground state, a gain in the potential energy may make the difference between energies in p- and s-states relatively small. The wave function of the p-state of a particle in a spherically symmetric potential is written as

$$\psi_{l,0}(r,\theta) = \sqrt{\frac{2l+1}{4\pi}} \frac{\chi_l(r)}{r} P_l(\cos\theta). \tag{13}$$

The radial part $\chi_l(r)$ of this wave function is required to satisfy the Schrödinger equation

$$-\frac{\hbar^2}{2m}\frac{\partial^2 \chi_l}{\partial r^2} + \left(\frac{\hbar^2}{2m}\frac{l(l+1)}{r^2} + U(r)\right)\chi_l = E_l \chi_l, \tag{14}$$

in which the potential energy $U(r)$ is given by (3) in the case of the bcc structure, $l=1$ is for p-state. To calculate the energy of the first excited state of helium atom in the helium medium, the direct variational method was used. The trial function $\chi_l(r)$ was chosen to be of the form

$$\chi_1(r) = C_1 r \sinh\left(\frac{\alpha r}{2\sqrt{3}}\right) e^{-s\cosh\left(\frac{\alpha r}{\sqrt{3}}\right)} \left[1+\gamma \sinh\left(\frac{\alpha r}{2\sqrt{3}}\right)\right]^{\beta} \qquad (15)$$

where $s, \gamma$ and $\beta$ are variational parameters, and $C_1$ is a normalization factor.

In the Fig.2, the ground state energy and the first excited state energy as functions of the medium expansion $b = \alpha(R - R_0)$ are presented. As seen from Fig.2, the energy gap between s and p bands of liquid $He^4$ has value ~10 K at the density of helium II ($\rho \approx 0.145 g/sm^3$).

## 3. Helium II

Possibility of a phase transition implies existence of internal parameters in a system, whose continuous or discontinuous change leads to a change of the phase state of the medium. In a neutral, simple one-component fluid, such as liquid $He^4$, there is the only internal parameter of the medium, that is the quantum state of the helium atom. At $T=0$, all the $N$ atoms $He^4$ are in the ground state.

As temperature increases, a part of atoms $n_p$ will pass into the $p$-state leading to an increase in the configurational entropy of the system. Moreover, the presence of helium atoms in different quantum states will lead to the quantum exchange interaction between neighboring $s$ and $p$ atoms which determines the delocalization of $s$ or $p$ atomic states over the system. Quantum delocalization of $s$ or $p$ atomic states over the system will lead to the stabilization of the interatomic distances in helium II, i. e. the density of liquid helium will be practically independent of temperature. The effect is similar to the stabilization of the internal medium parameter by the quantum particle of the electron, which is well known in the case of polarons [8] or fluctuons [9].

## 4. Conclusion

Existence of a quantum discrete spectrum of excitations of helium atoms is the main cause of the formation of a gap $\Delta\varepsilon_p$ in the spectrum of a helium system and, as a consequence, the superfluidity of liquid helium. Indeed, at $T=0$, all the helium atoms in liquid helium are in the ground state.

Because of presence of the gap in the energy spectrum of the liquid, atoms are not excited at low velocities of the fluid $\mathbf{v}$ ($\Delta\varepsilon_p > |\mathbf{pv}|$), and fluid flow occurs without the viscous friction, i.e. in a superfluid way. If the flow velocity exceeds a certain critical value, the energy of atoms exceeds the gap and the property of superfluidity disappears. The excitation of atoms in a quasi-continuous zone may also occur with increasing temperature, in this case, a part of the helium atoms may remain in the ground superfluidity state, i.e. the system will be in the two-fluid state. The mechanism of supurfluidity under consideration is very similar to the physical mechanism of the classical superconductivity in which the superfluidity of electronic liquid is determined by presence of a gap in electronic spectrum.

We can consider the effects of electric induction [3] and the resonant absorption of microwaves [4] observed in superfluid $He^4$ as an indirect proof of presence of quantum transitions in $He^4$. These effects can be associated with violation of the adiabatic condition in electron-nuclear subsystem $He^4$ in the course of quantum transition of helium atoms.

The quantum $s-p$ transition of an helium atom leads to a change in the distribution of positions of the atom in a potential well that leads to the dramatic changes of the character of motion of the nucleus and electron subsystem of the helium atom, and initiates of a short-time dipole moment of an atom.